\documentclass[aps,prl,superscriptaddress,twocolumn,showpacs]{revtex4-1}
\usepackage{graphicx}
\usepackage{amssymb,amsmath}
\begin{document}

\title{Coherent spin control by electrical manipulation of the magnetic anisotropy}
\author{Richard E George}
\affiliation{Department of Materials, University of Oxford, Parks Road, Oxford OX1 3PH}
\affiliation{CAESR, Clarendon Laboratory, Department of Physics, Parks Road, Oxford, OX1 3PU}
\author{James P Edwards}
\affiliation{CAESR, Clarendon Laboratory, Department of Physics, Parks Road, Oxford, OX1 3PU}
\author{Arzhang Ardavan}
\affiliation{CAESR, Clarendon Laboratory, Department of Physics, Parks Road, Oxford, OX1 3PU}
\date{\today}
\begin{abstract}
High-spin paramagnetic manganese defects in polar piezoelectric zinc oxide exhibit a simple almost axial anisotropy and phase coherence times of the order of a millisecond at low temperatures. The anisotropy energy is tunable using an externally applied electric field. This can be used to control electrically the phase of spin superpositions and to drive spin transitions with resonant microwave electric fields.
\end{abstract}
\maketitle

Couplings between magnetic and electric degrees of freedom give rise to such fundamental phenomena in condensed matter as multiferroelectricity~\cite{Khomskii:2009p4297,SpaldinScience2005}, unconventional superconductivity~\cite{Monthoux:2007p4329} and spin-density-waves~\cite{Gruner1994}, and are key to proposed future technologies such as high sensitivity metrology~\cite{Zhai:2006p4831} and quantum-dot-based quantum information processing (QIP)~\cite{Petta:2005p4568}. The control of spin states using electric fields rather than magnetic fields is particularly valuable for QIP~\cite{Bassett:2011je}, because electric fields can be applied on short length scales, and down to the scale of qubit separations in a device~\cite{Andre:2006p4760,Bernien:2012hy,Ochsenbein:2011hl}. In this Letter we identify a class of electrically-controllable spin qubits: high-spin magnetic defects in polar semiconductor hosts. In one member of this class, the manganese substitutional defect in a crystalline zinc oxide host, we demonstrate coherent oscillations of the spin state driven by d.c.\ electric field pulses and spin transitions driven by resonant microwave electric fields.

Electrical control has been considered in various candidate physical systems for spin qubits, both experimentally and theoretically. In lithographically defined GaAs quantum dots, spin-orbit coupling permits coherent control of an electron spin by applying a resonant high frequency voltage to one of the confining gates~\cite{Nowack:2007du}. Stark shifts in spin-splittings, originating from hybridization with excited state orbitals, have been observed in self-assembled quantum dots~\cite{Smith:2005p4722} and in bound donors in semiconductors~\cite{Bradbury:2006p1699,Morello:2010p4597} and in defects in diamond~\cite{vanOort:1990vq}, while frustrated spin triangles exhibit spin-electric couplings via modulation of exchange interactions~\cite{Trif:2008PRL,Trif:2010p3957}. Here, our approach is to manipulate the electrostatic environment of a high-spin paramagnetic defect, manganese, in a polar piezoelectric host material, zinc oxide, thereby controlling electrically the spin Hamiltonian of the defect.
 
\begin{figure*}
\begin{center}
\includegraphics[width=\textwidth]{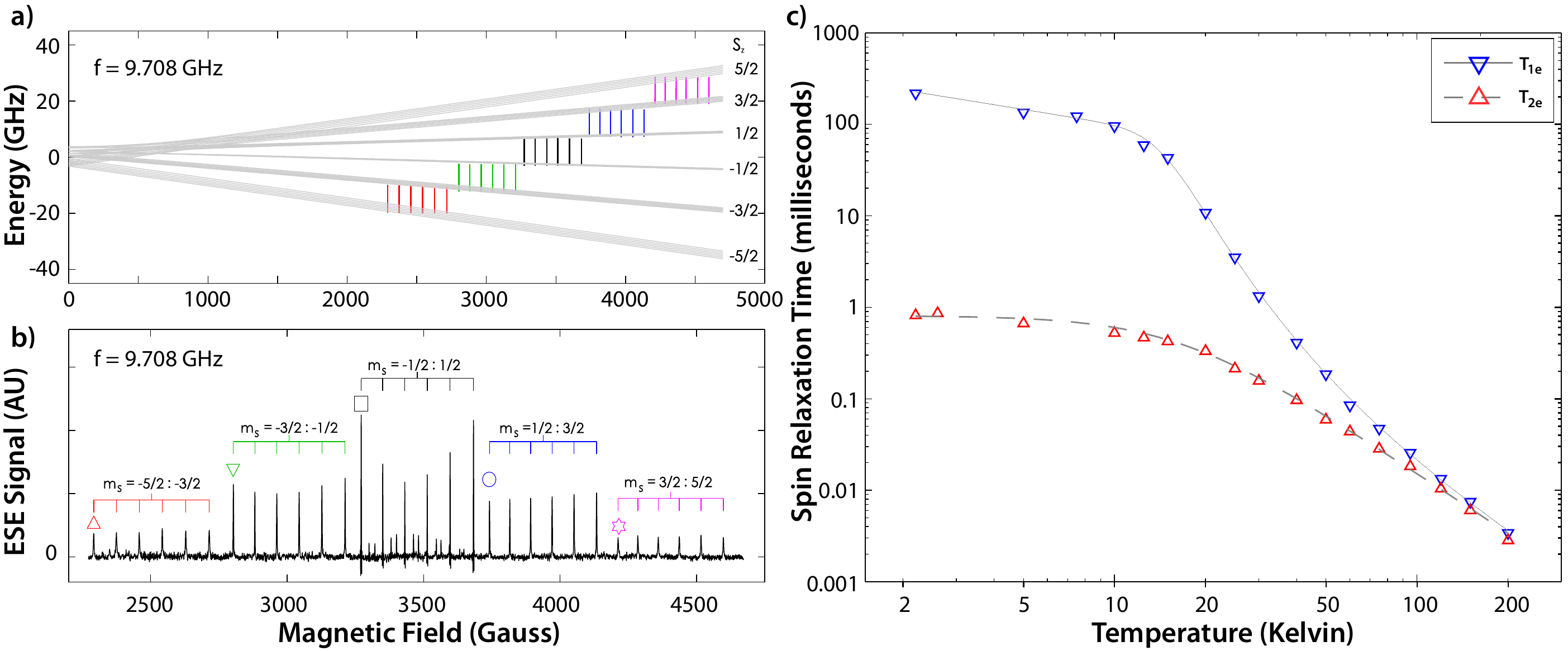}
\end{center}
\caption{\label{fig1} 
(Color online)
{(a)} The energy level structure of the system as a function of magnetic field.
{(b)} The ESR spectrum comprises five groups ($\Delta m_s = \pm 1$) of six (2I+1) lines. Weakly allowed $(\Delta m_s,\Delta m_I) =(\pm1,\pm1)$ transitions are visible near the centre of the spectrum. 
\textbf{c)} Electron spin lifetimes $T_{1e}$ and $T_{2e}$ for the $m_s = -1/2:1/2$ transition. Fits are to $T_{1e} = A T^2 / ( \exp( -  E / k_B T) + 1)$ modelling spin relaxation via phonons, $T_{2e} \sim {(T_{1e})}^{2/3}$ modelling relaxation caused by spin flips of neighbouring Mn spins. 
}
\end{figure*}

The spin Hamiltonian of manganese defects in zinc oxide (ZnO:Mn) is well-known from continuous-wave ESR~\cite{Hausmann:1968p1340}. The polar environment lifts the spin degeneracy of the $S=5/2$ Mn$^{2+}$ ions which are substitutional on Zn$^{2+}$ sites, while the $I=5/2$ nuclear spin of $^{55}$Mn is coupled to the electron spin by the hyperfine interaction. The Hamiltonian, including a magnetic field which lifts the remaining Kramer's degeneracies is
\begin{equation}
H = \mu_B \hat{\mathbf{S}} \cdot \mathbf{g} \cdot \mathbf{B} + g_N \mu_N \hat{\mathbf{I}} \cdot \mathbf{B} + \mathbf{\hat{S}} \cdot \mathbf{A} \cdot \hat{\mathbf{I}} + H_{\mathrm ZFS} 
\label{hamiltonian}
\end{equation}
where the axial anisotropy, or zero field splitting, term
\begin{equation}
H_{\mathrm ZFS} = - D \hat{S}_z^2 .
\end{equation}
Higher order effects, such as fourth order electron spin anisotropy and nuclear quadrupole terms, have been detected~\cite{Hausmann:1968p1340}, but they are small and not relevant for the following discussion. The resulting energy spectrum is shown as a function of magnetic field applied at a small angle to the anisotropy axis in Fig.~\ref{fig1}(a). The absorption in response to an oscillating magnetic field of frequency 9.7~GHz, applied perpendicular to the static field, is shown in Fig.~\ref{fig1}(b); the peaks are magnetic dipole electron spin resonance transitions corresponding to the selection rule $\Delta m_s = \pm 1$.
 
The spin states are highly coherent. Fig.~1(c) shows the relaxation times for electron spin population ($T_{1e}$) and coherence ($T_{2e}$) as a function of temperature, measured on the $m_s=\pm 1/2$ transition 
($\square$ symbols in Fig~1b). 
The relaxation times increase rapidly below room temperature with $T_{1e} \sim T^{-3}$ and $T_{2e} \sim T^{-2}$, saturating at $T_{1e} \approx 0.2$~s and $T_{2e} \approx 0.8$~ms below 10~K. The cubic temperature dependence of $T_{1e}$ is consistent with scattering by a population of phonons, an indication that lattice distortions couple to the Mn spin.

The total Mn density is $4.0 \pm 0.1 \times 10^{16}$~cm$^{-3}$, corresponding to a mean separation of 17~nm between nearest neighbours. The thermal spin populations lead to an effective dilution of on-resonance spins by a factor of $(2S+1)(2I+1)=36$, so spin centres in the same state are separated on average by 58~nm. The quadratic $T_{2e}$ temperature dependence is characteristic of decoherence driven by spin flips (i.e. $T_{1e}$ relaxation) of off-resonant neighbouring spins~\cite{Barbon:1999p4615,Tyryshkin:2011fi}, and the saturation below 10~K is consistent with magnetic dipole couplings between more distant spins in the same sub-level. 

The relaxation times are independent of the nuclear spin projection to within experimental errors, but as expected for a phonon-driven mechanism, $T_{2e}$ depends on the $m_s$ states involved in the superposition. The coherence times for $m_s = \pm 1/2: \pm 3/2$ superpositions and $m_s = \pm 3/2: \pm 5/2$ are, respectively, about 0.65 and 0.58 times that for $m_s = -1/2 : +1/2$. The ratios are not strongly temperature-dependent.

The large moment of the $^6$S$_{5/2}$ electronic configuration allows us to perform Rabi oscillations with periods down to $\sim 5$~ns (depending on the $m_s$ states involved) with standard apparatus, allowing us to perform $\sim 4 \times 10^{6}$ single qubit operations within the relaxation time of this system.
 
\begin{figure}
\begin{center}
\includegraphics[width=\columnwidth]{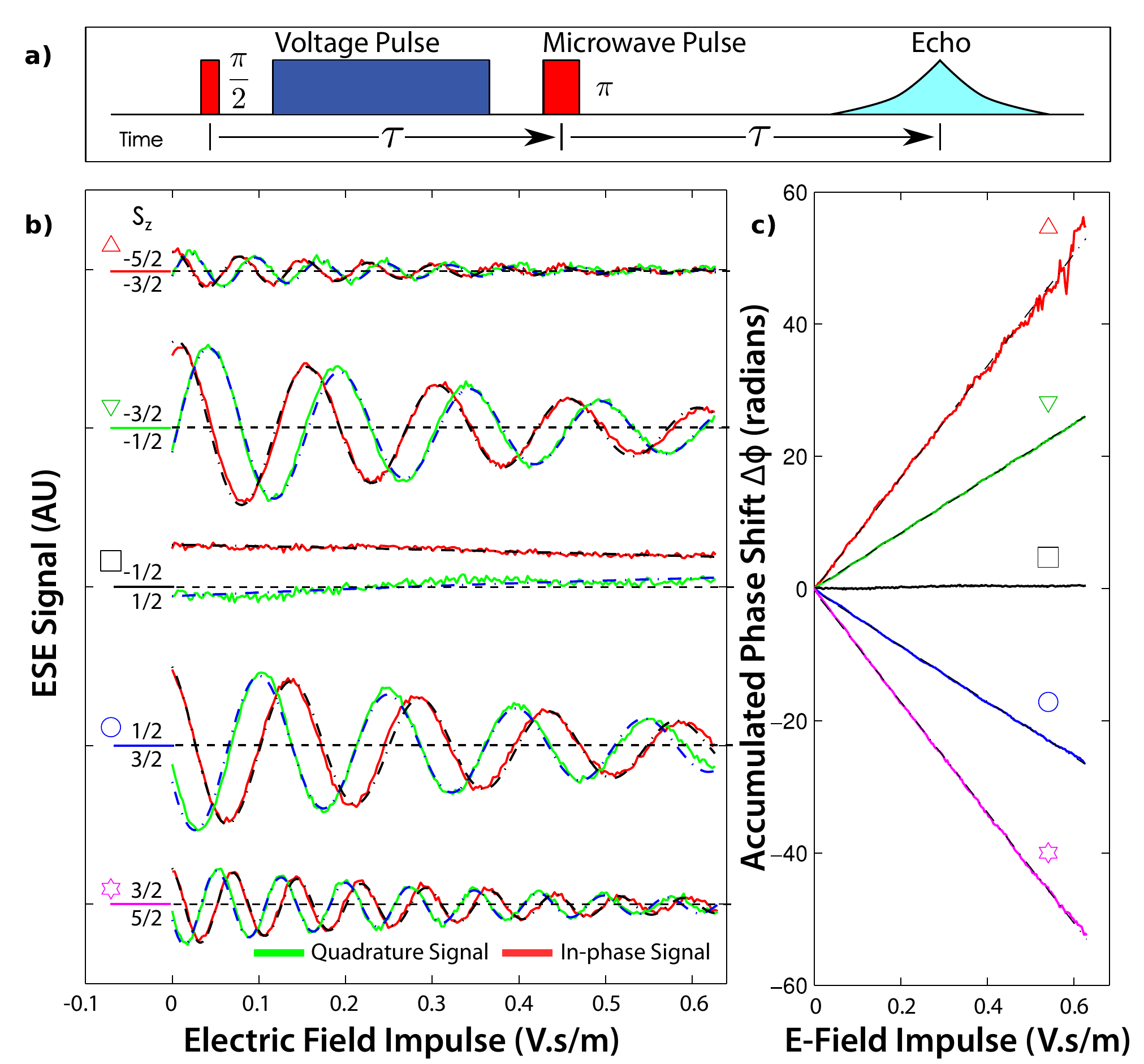}
\end{center}
\caption{\label{fig2} 
(Color online)
{(a)} A Hahn echo sequence enclosing a voltage pulse to generate phase shifts of the spin coherence.
{(b)} The phase of the recorded spin echo shifts when an electric field impulse is applied. For the $m_s < 0$ transitions the electric field generates a positive phase shift (i.e.\ the quadrature channel $Q$, plotted in green, lags the in-phase channel, $I$, red), whereas for $m_s > 0$ a negative phase shift occurs. {(c)} The recorded oscillations are unwrapped using $\Delta\phi = \tan^{-1}(Q / I)$ and plotted against the electric field impulse. The phase shift is proportional to the electric field impulse, indicating that the shift in $D$ is linear in $E$. }
\end{figure}

The origin of the axial anisotropy of the Mn spins, parameterised by $D$ in Eqn.~\ref{hamiltonian}, is the electric field experienced by the Mn defects in the polar ZnO. Applying an additional external electric field $E$ along the polar (100) axis allows us to modify the zero-field-splitting parameter such that
$D = D_0  + \kappa E $,
and it is possible to manipulate the spin via this externally-controllable parameter. Fig~\ref{fig2}(a) shows spin state control using a d.c.\ electric field pulse embedded into a Hahn echo sequence~\cite{MimsLEFE1976,Bradbury:2006p1699}. The initial $\pi/2$ pulse generates a superposition between adjacent $m_s$ levels; the electric field adjusts their splitting for a time $\tau_p$, causing a phase 
\begin{equation}
\Delta \phi = \alpha \kappa  E \tau_p 
\label{phase_shift}
\end{equation}
to accumulate in the superposition; the $\pi$-pulse refocuses static inhomogeneities; and an echo forms whose phase reflects the manipulation applied by the electric field pulse. Phase shifts in quantum spin coherences,
generated by sweeping the d.c.\ electric field $E$ between 0 and $4 \times 10^5$~V/m for a pulse duration of $\tau_p=6~\mu$s, are converted to populations and are shown in Fig.~2(b) for the five coherences between adjacent $m_s$ states. The electric-field-induced shifts in energy are proportional to $m_s^2$, so the $m_s = \pm 5/2: \pm 3/2$ superpositions respond most strongly to $E$ ($\alpha = 4$ in Eqn.~\ref{phase_shift}), the $m_s = \pm 3/2: \pm 1/2$ respond more weakly ($\alpha = 2$), and the $\pm 1/2$ superposition is unaffected ($\alpha=0$) to first order. We find only a first order phase shift from the applied electric field~\cite{Bradbury:2006p1699}. 
The rate of accumulation of phase gives a direct measurement of the coupling to the electric field,  $\kappa = 20.8\pm0.2$~rad\,s$^{-1}$/V\,m$^{-1}$.

The oscillations in Fig.~\ref{fig2}(b) constitute coherent manipulations of the Mn spin using an electric field. The magnitude of the effect that we observe suggests that the externally applied electric field does not simply add to the internal electric field of the ZnO. Assuming that the zero-field splitting in the absence of an applied electric field, 
$D_0= 1.585\pm0.006 \times 10^9$~rad\,s$^{-1}$, 
originates from an internal electric field of a typical magnitude, $\sim 10^{10}$~V\,m$^{-1}$, the magnitude of $\kappa$ that we measure demonstrates that the external applied electric field is several orders of magnitude more effective at modifying $D$ than the internal electric field. This is probably because the dielectric and piezoelectric response of the ZnO contributes to the modification of the local environment of the Mn ion, thereby enhancing the effect of the applied electric field on the spin anisotropy. While demonstrated here for Mn defects in ZnO, this effect should be observable for any $S \ge 1$ spin in a lattice whose crystal-field is modified by an external electric field.

The capability to tune spins into and out of resonance with a cavity using an electric field is highly desirable for information processing applications~\cite{Andre:2006p4760}. Electric fields of order $10^6$~V\,m$^{-1}$, that might be applied by microscopic lithographically-defined structures, would generate shifts in the spin resonance frequencies of order 10~MHz. Comparing this tuning bandwidth to the $\sim$~100~kHz linewidths of contemporary superconducting microwave resonators~\cite{Leek:2010p4927}, we can envisage hybrid systems in which d.c.\ electrical tuning via an array of surface gates selectively brings chosen spin qubits into and out of resonance with a superconducting cavity bus.
 
\begin{figure}
\begin{center}
\includegraphics[width=\columnwidth]{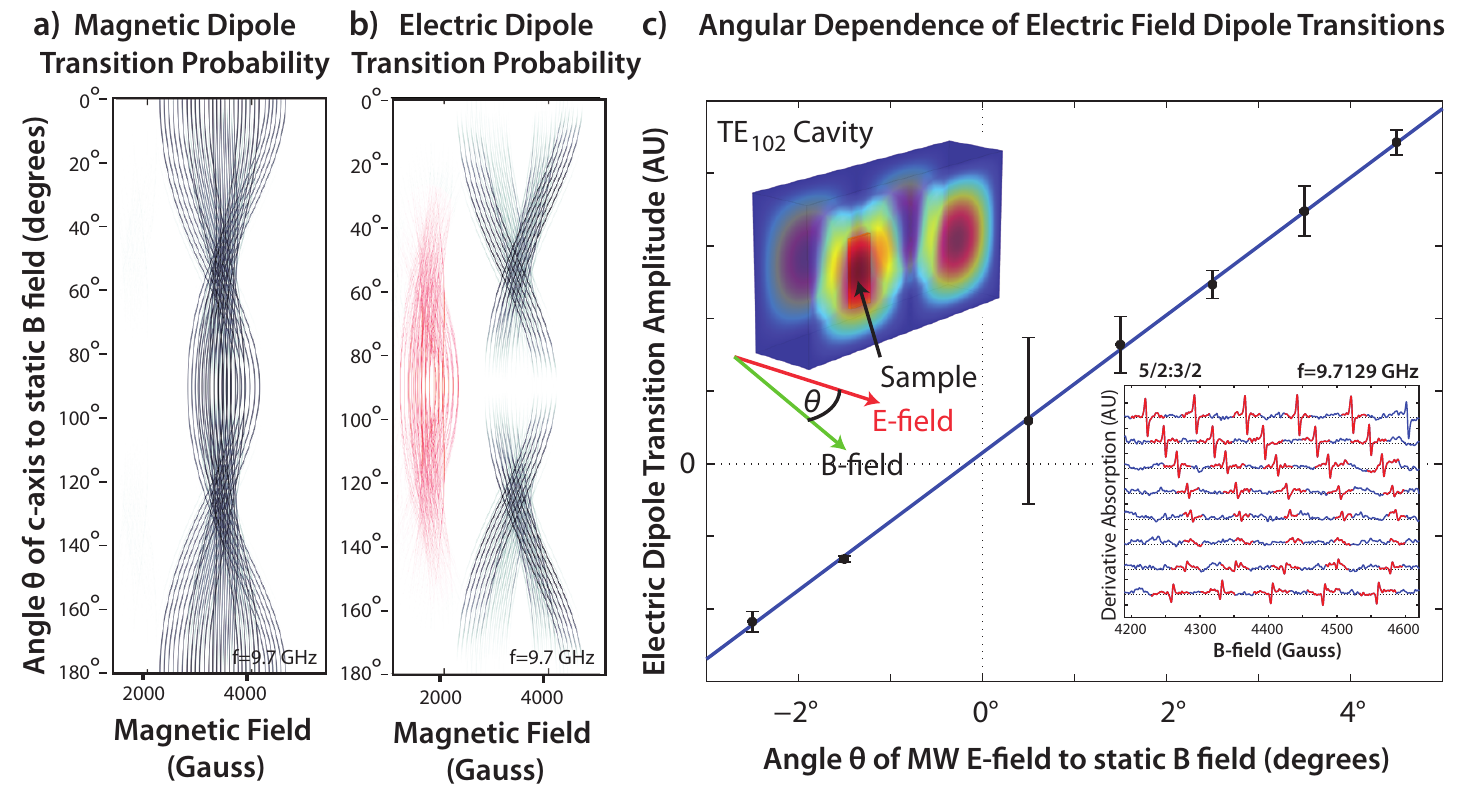}
\end{center}
\caption{\label{fig4} 
(Color online)
{(a)} (Theoretical) Electric dipole spin transition probability per unit time (indicated by colour density) in ZnO:Mn as a function of the angle $\theta$ between the static magnetic field and the anisotropy axis for an oscillatory electric field of frequency 9.7~GHz. Single quantum coherences are plotted in black, double quantum coherences in red. 
{(b)} (Theoretical) The corresponding conventional magnetic dipole electron spin resonance transition probability, for an oscillating magnetic field of the same frequency applied perpendicular to the static magnetic field.
{(c)} (Experimental) The measured electric dipole transition strengths for small angles $\theta$. The upper inset shows the experimental geometry; the oscillatory electric field strength in the rectangular $TE_{102}$ mode is indicated by colour density (red is strong, blue is weak). The lower inset shows the raw data for the derivative absorption spectra as a function of magnetic field for small angles $\theta$.
}
\end{figure}

We have seen so far how d.c.\ electric fields can shift energy levels and cause phases to accumulate in preexisting spin state superpositions. However, it is also possible to create superpositions and excite coherences using resonant a.c.\ electric fields. For a magnetic field applied along $z$ at an angle $\theta$ to the anisotropy axis, the zero field splitting Hamiltonian becomes 
\begin{equation}
H_{\mathrm{ZFS}} ( \theta ) = -D ( \hat{S}_z \cos \theta + \hat{S}_x \sin \theta)^2 .
\label{crystal_field_with_angle}
\end{equation}
If $D$ is modulated at high frequency by applying an oscillatory electric field along the anisotropy axis, $H_{\mathrm{ZFS}}$ can generate resonant transitions between spin states $m_s$. Fig.~\ref{fig4}(a) shows the transition strengths driven by an electric field of frequency 9.7~GHz as a function of magnetic field strength and orientation $\theta$. When the magnetic field is perpendicular to the anisotropy axis and the oscillating electric field ($\theta = 90^\circ$), $H_{\mathrm{ZFS}} = -D \hat{S}_x^2 = - \frac{1}{4}D( \hat{S}_{+} + \hat{S}_{-})^2$. $H_{\mathrm{ZFS}}$ contains pairs of raising and lowering operators, so double quantum coherences may be generated by a resonant oscillating electric field, i.e.\ transitions can be excited for which $\Delta m_s = \pm 2$; these transitions are shown in red in Fig.~\ref{fig4}(a). For a general non-zero $\theta$, $H_{\mathrm{ZFS}}$ contains first and second powers of the raising and lowering operators, driving both single and double quantum coherences (transitions for which $\Delta m_s = \pm 1$ or $\pm 2$). For $\theta=0$, when both fields are parallel to the anisotropy axis, the total spin Hamiltonian is secular and no transitions are excited. For reference, the conventional electron spin resonance magnetic dipole transitions are shown in Fig.~\ref{fig4}(b), for a static magnetic field applied at an angle $\theta$ to the anisotropy axis and an oscillatory magnetic field applied perpendicular to the static magnetic field.

Fig.~\ref{fig4}(c) confirms that we can excite electric dipole transitions between spin states. The sample is positioned on the conducting wall in an electric field antinode (amplitude $\sim500$~V/cm) of a rectangular TE$_{102}$ resonator ($Q \sim 5000$), such that the oscillating electric field (which is perpendicular to the conducting wall) is parallel to the anisotropy axis. The orientation of the static magnetic field, which for $\theta=0$ is parallel to the oscillatory electric field, can be varied in our apparatus by up to about 5~degrees. The inset shows the derivative absorption spectrum obtained as a function of magnetic field strength as $\theta$ is varied; the magnetic field range is chosen to examine the single quantum coherence transitions $m_s = +5/2: +3/2$ because, along with $m_s = -5/2: -3/2$, these should exhibit the strongest electric dipole transitions. For $\theta = 0$ no transitions are visible, and as the magnetic field is rotated away from the anisotropy axis, the transition strengths increase in proportion to the component of the oscillating electric field perpendicular to the magnetic field. This is exactly the behaviour expected for the electric dipole transitions as predicted in Fig.~\ref{fig4}(a). 

We can rule out the possibility that the transitions in Fig.~\ref{fig4}(c) are magnetic dipole transitions. Firstly, the sample is small compared to the cavity dimensions, so the oscillating magnetic field that it experiences is vanishingly small. Secondly, any such oscillating magnetic field that the sample does experience is parallel to the conducting wall and is therefore perpendicular to the static magnetic field for $\theta = 0$; in this geometry the magnetic dipole transitions are maximised and, to first order, insensitive to small changes in $\theta$.

We have shown that the electron spins of substitutional manganese in zinc oxide exhibit very long coherence times and that their states may be manipulated rapidly using d.c.\ and a.c.\ electric fields. Owing to the possibility of applying electric fields over much shorter length scales than magnetic fields, such electric field control of qubits offers dramatic advantages for quantum information processor architectures~\cite{Wu:2010fm,Kubo:1269416,Zhu:1401583}. 
The spin-electric coupling is several orders of magnitude larger than is found in other electric-field-sensitive spin qubit candidates~\cite{vanOort:1990vq,Bradbury:2006p1699}, and large enough that electrical control of the spin state is possible on time scales comparable with magnetic control: the coupling of the Mn$^{2+}$ spin to the electric field of a plane wave in bulk ZnO is about one order of magnitude smaller than the strength of its coupling to the magnetic field; the electric field coupling could be enhanced by confining the electromagnetic wave in a structure with a higher characteristic impedence than the bulk.
The origin of the electric field coupling that we identify here should be applicable to many high-spin magnetic defects in polar materials. In particular, we would expect that electrically charged Fe$^{3+}$ impurities in ZnO should show even larger couplings than Mn$^{2+}$ impurities, which are isoelectronic with the Zn$^{2+}$. Other polar materials, such as the ferroelectric strontium titanate, should also be expected to host magnetic defects exhibiting strong electric field couplings.

We thank W.~Hayes, A.~Heinrich, T.~Lancaster and J.~Moeller for helpful discussions. We thank J.J.L.~Morton for discussions and materials. This research was supported by the UK EPSRC. A.A.\ thanks the Royal Society for support. 


%

\end{document}